\documentclass[aps,prd,amsfonts,preprint,
eqsecnum,nofootinbib]
{revtex4}
\usepackage{graphics,epsfig}
\begin{document}
\preprint{WM-02-106}
%
\title{\vspace*{0.3in}
Noncommutative Gauge Theory without Lorentz Violation
\vskip 0.1in}
\author{Carl E. Carlson}
\email[]{carlson@physics.wm.edu}
\author{Christopher D. Carone}
\email[]{carone@physics.wm.edu}
\affiliation{Nuclear and Particle Theory Group, Department of
Physics, College of William and Mary, Williamsburg, VA 23187-8795}
\author{Nahum Zobin}
\email[]{zobin@math.wm.edu}

 \affiliation{Department of Mathematics, College of William and Mary, 
Williamsburg, VA 23187-8795}

\date{June 4, 2002}
\begin{abstract}
The most popular noncommutative field theories are characterized by a 
matrix parameter $\theta^{\mu\nu}$ that violates Lorentz invariance. We 
consider the simplest algebra in which the $\theta$-parameter is promoted 
to an operator and Lorentz invariance is preserved.  This algebra arises
through the contraction of a larger one for which explicit representations
are already known. We formulate a star product and construct the 
gauge-invariant Lagrangian for Lorentz-conserving noncommutative QED.  
Three-photon vertices are absent in the theory, while a four-photon 
coupling exists and leads to a distinctive phenomenology.
\end{abstract}
\maketitle


\section{Introduction}\label{sec:intro}

Over the past few years, the thrust of ``beyond the standard model''
particle theory has undergone a fundamental shift, from exploration
of extensions of the standard model in flat, four-dimensional spacetime
to those that follow from modifications of the structure of spacetime
itself.  One such possibility is the existence of extra spatial dimensions 
with either large or infinite radii of compactification, 
an idea motivated by the desire to eliminate the hierarchy between the 
gravitational and the weak scale.  Aside from the existence of extra 
dimensions themselves, the reduction in the fundamental scale in these 
scenarios opens the possibility that new phenomena arising in 
string theory may also become of experimental relevance.  One fascinating 
possibility that has met considerable interest in the recent literature
is that spacetime may become noncommutative at distance scales just below 
those currently accessible in 
experiments~\cite{HK,HPR,mncqed,CHK,MPR,ABDG,CCL,ncxd,GMW,posp2,iltan,ACDNS,h2}. 
In the canonical version of noncommutative spacetime, the position four-vector $x^\mu$ 
is promoted to an operator satisfying the commutation relation
\begin{equation}
[\hat{x}^\mu \, , \, \hat{x}^\nu] = i \, \theta^{\mu\nu} \,\,\, ,
\label{eq:canonical}
\end{equation}
where $\theta^{\mu\nu}$ is a real, constant matrix of ordinary c-numbers.
Precisely this situation is realized in string theory when open
strings propagate in the presence of a constant background
antisymmetric tensor field~\cite{SW}.  Keeping in mind that all scales in 
nature may not be far above the weak scale, it is not unreasonable to
consider the possibility that Eq.~(\ref{eq:canonical}) could lead to 
observable consequences.

Connecting Eq.~(\ref{eq:canonical}) to experimental observables requires
that one formulate quantum field theories on a noncommutative 
space~\cite{Madore,HAY,ARM,LIA,CSJT,grm,Jurco,Calmet}.
While ordinary fields are functions of a commuting, classical position
four-vector $x^\mu$, the algebraic properties of the underlying noncommutative
theory can be reproduced by replacing ordinary multiplication by a
star product.  For example, in the canonical case, one defines 
a mapping between functions of noncommuting coordinates $\hat{x}^\mu$ 
and functions of the c-number coordinates $x^\mu$ via the Fourier transform
\begin{equation}
\hat{f}(\hat{x}) = \frac{1}{(2\pi)^n} \int d^n k\,  e^{-i k \hat{x}} \int d^n x
\,  e^{ikx} f(x)  \,\,\,.
\end{equation}
The requirement that
\begin{equation}
\hat{f}\hat{g}= \widehat{f\star g} \,\,\, ,
\label{eq:thereq}
\end{equation}
{\em i.e.} that the functions $f(x)$ and $g(x)$ yield a representation of
the algebra under star multiplication, allows one to define the star
product.  In the canonical case, one obtains the Moyal-Weyl result:
\begin{equation}
(f \star g)(x) = f(x)\, \exp[\frac{i}{2} \stackrel{\leftarrow}
{\partial_\mu} \theta^{\mu\nu} \stackrel{\rightarrow}{\partial_\nu}]\,
g(x) \,\,\, .
\label{eq:moyal}
\end{equation}
A field theory action can now be represented as a functional of fields that 
depend only on commuting spacetime coordinates
\begin{equation}
S = \int d^4 x \, {\cal L}(\phi(x),
\partial_\mu \phi(x))_\star \,\,\, ,
\label{eq:acform}
\end{equation}
where the $\star$ subscript indicates that all multiplications between
fields are defined by Eq.~(\ref{eq:moyal}).  This representation of the action is 
nothing more than the mapping of the operator trace
\begin{equation}
S = \mbox{Tr } \hat{{\cal L}}
\label{eq:actdef}
\end{equation}
to the space of ordinary functions.

Formulating gauge theories on noncommutative spaces introduces additional
complications.  For example, the simplest formulation of noncommutative
U(1) gauge theory (one that does not require working order by order
in the parameter $\theta$) is only consistent if matter fields have
charges $0$ or $\pm 1$; adding additional states with other charges 
makes it impossible to define a covariant derivative~\cite{HAY}.  While U(N) 
gauge theories follow with relatively little effort after promoting  ordinary
to star multiplication~\cite{ARM}, SU(N) gauge theories cannot be constructed in
such a straightforward manner. These problems have been surmounted by
Jur\v{c}o, {\em et al.}~\cite{Jurco}, who have shown that it is possible to 
maintain gauge invariance and noncommutativity simultaneously by employing 
a nonlinear field redefinition that is determined order by order in 
an expansion in the parameter $\theta$.  This approach has allowed 
construction of the full noncommutative standard model~\cite{Calmet,lvncsm}, without 
relying on awkward embeddings of the standard model gauge group.

The most notable phenomenological feature of canonical noncommutative
field theories is the violation of Lorentz invariance following from
Eq.~(\ref{eq:canonical})~\cite{HPR,mncqed,CHK,MPR,ABDG,CCL}.  Both $\theta^{i0}$ 
and  $\epsilon^{ijk} \theta_{jk}$ are fixed three-vectors that define
preferred directions in a given Lorentz frame.  Phenomena such as
the diurnal variation of collider cross sections have been noted in 
studies of noncommutative QED~\cite{HPR,mncqed}, 
even though some bounds~\cite{MPR,ABDG,CCL} from low-energy tests of Lorentz 
invariance seems to suggest that effects at colliders are likely to be negligible. 
Such constraints have been shown to be even more significant in noncommutative 
QCD~\cite{CCL}, and are likely to persist in more general canonical models.

One approach to this problem is to ignore it, on the grounds that (most of)
the bounds in question are obtained in theories whose Lagrangians are known
only at lowest order in $\theta$.  These theories have not been shown to
be renormalizable,  while the most dangerous effects are obtained only 
through loop corrections~\cite{ABDG,CCL}.  On the other hand, in simple situations where 
both all-orders and lowest-order Lagrangians are known, the bounds on 
Lorentz violation from loop effects are even {\em stronger} in the full 
theory~\cite{ABDG}.  We take the position that low-energy tests of Lorentz 
invariance~\cite{lv} are likely to present a generic impediment to formulating 
a noncommutative standard model that is based on the canonical relation 
Eq.~(\ref{eq:canonical}) and that is also phenomenologically relevant.  
One alternative is to push the noncommutativity into extra 
dimensions~\cite{ncxd,GMW,posp2}, leaving the four ordinary space-time dimensions 
commutative and Lorentz invariant~\cite{ncxd,GMW}.   This has added benefits, for 
example, in allowing one to formulate a simple noncommutative QED including matter 
fields with arbitrary charges, provided these fields are restricted to an orbifold 
fixed point~\cite{ncxd}.  A more challenging approach is to formulate noncommutative 
field theories that are free from Lorentz violating effects, {\em ab initio}.  It is
this  approach we wish to explore in our present work.

In this paper we will consider a new class of noncommutative theories in
which the parameter $\theta$ in Eq.~(\ref{eq:canonical}) is promoted to an 
operator $\hat{\theta}^{\mu\nu}$  that is {\em not} constant, but transforms
as a Lorentz tensor.  In the next section, we show that this
algebra can be interpreted as a contraction of a famous Lorentz-invariant 
algebra due to Snyder~\cite{snyder} for which explicit representations are 
known.  By treating $\hat{\theta}$ as an unphysical parameter, we find
the appropriate generalizations of the star product and operator trace for
functions of both $x^\mu$ and $\theta^{\mu\nu}$.  We then show how these
results may be applied in constructing Lorentz-invariant Lagrangians for
fields that are functions of $x^\mu$ alone.  In the case of gauge theories,
we accomplish this last step using the type of nonlinear field redefinitions 
introduced in the context of noncommutative SU(N) gauge theories~\cite{Jurco}.
As a concrete example, we formulate Lorentz-invariant noncommutative QED and 
show that 4-photon interactions are present, while vertices with an odd 
number of photons do not occur.  In the fourth section we undertake a brief 
phenomenological investigation of light-by-light scattering in this theory, 
and in the final section we summarize our conclusions.


\section{Algebra and Star-Product}\label{sec:algebra}


Let us consider the simplest generalization of Eq.~(\ref{eq:canonical})
in which $\theta^{\mu\nu}$ is promoted to an operator $\hat \theta^{\mu\nu}$ 
in the same algebra as the coordinates:
\begin{eqnarray}       \label{algebra}
\left[ \hat x^\mu , \hat x^\nu \right] &=& i \hat\theta^{\mu\nu} \,\, ,
   \nonumber \\
\left[ \hat \theta^{\mu\nu} , \hat x^\lambda \right] &=& 0 \,\, ,
   \nonumber \\
\left[ \hat \theta^{\mu\nu} , \hat \theta^{\alpha\beta} \right] &=& 0
  \ .
\label{eq:algebra}
\end{eqnarray}
One could proceed immediately to discuss the algebra of functions
$\hat{f}(\hat{x}, \hat{\theta})$, as well their mapping to ordinary 
functions $f(x,\theta)$ and the associated star product.  However, it is 
useful first to display an explicit representation of the operators
$\hat{x}$ and $\hat{\theta}$ that makes the Lorentz invariance of
Eq.~(\ref{eq:algebra}) manifest.  We accomplish this by contracting another 
Lorentz-invariant algebra for which representations are already known.

Snyder proposed an algebra of noncommutative spacetime coordinates
leading to a Lorentz-invariant discrete spacetime~\cite{snyder},  
\begin{eqnarray}
\left[ \hat x^\mu , \hat x^\nu \right] &=& 
    i {a^2} \hat M^{\mu\nu}  \,\, ,
                                             \nonumber \\
\left[ \hat M^{\mu\nu} , \hat x^\lambda \right] &=& 
    i \left( \hat x^\mu g^{\nu\lambda}
                   - \hat x^\nu g^{\mu\lambda} \right) \,\, ,
                                            \nonumber \\
\left[ \hat M^{\mu\nu} , \hat M^{\alpha\beta} \right] &=& 
    i  \left( \hat M^{\mu\beta} g^{\nu\alpha}
          +   \hat M^{\nu\alpha} g^{\mu\beta}
          -   \hat M^{\mu\alpha} g^{\nu\beta}
          -   \hat M^{\nu\beta} g^{\mu\alpha}     \right)
                                            \ ,
\end{eqnarray}
where $g^{\mu\nu}=diag\,(+,-,-,-)$. The last two commutation relations
involving the $\hat M^{\mu\nu}$  are those of the generators of the Lorentz group,
while the first is new~\cite{snyder,dop}.  (Together they imply
$\hat{M}$ and $\hat{x}/a$ can be identified as the generators of SO(4,1).)  Snyder's representation of 
this algebra is obtained by considering  a 5-dimensional space with coordinates $\eta_0$,
\ldots, $\eta_4$ and  metric $diag\,(+,-,-,-,-)$, on which ordinary Lorentz
transformations act  only on the first four coordinates.  Let us define
$\eta_\mu \equiv (\eta_0, \eta_1, \eta_2, \eta_3)$,
$\eta^\mu \equiv (\eta_0, -\eta_1, -\eta_2, -\eta_3)$,  and
\begin{eqnarray}
\hat x^\mu &=& i a \left( \eta_4 \frac{\partial}{\partial \eta_\mu} + 
\eta^\mu \frac{\partial}{\partial \eta_4}
      \right)  \,\, ,
                           \nonumber \\
\hat M^{\mu\nu} &=& i  
  \left(  \eta^\mu \frac{\partial}{\partial \eta_\nu} - \eta^\nu 
\frac{\partial}{\partial \eta_\mu}  \right)
                            \ .
\label{eq:therep}
\end{eqnarray}
Transformations that leave both $\eta_4$ and the
quadratic form $\eta_0^2 - \eta_1^2 - \eta_2^2 - \eta_3^2 - \eta_4^2$ 
invariant are Lorentz transformations of the $\eta_\mu$; such transformations
induce ordinary Lorentz transformations on the coordinates $\hat x^\mu$.  
From Eq.~(\ref{eq:therep}), it is not hard to show that the spatial coordinate 
operators $\hat{x}^i$ do not have a continuous spectrum, but rather have 
eigenvalues that are integers times the length scale $a$.  The time 
coordinate $x^0$, on the other hand, can be shown to have a continuous spectrum.

The contraction of an algebra is a simpler one obtained by taking the limit 
of some parameter.  We consider the rescaling
\begin{equation}
\hat M^{\mu\nu} =  \hat \theta^{\mu\nu} / b \ .
\end{equation}
and the limit 
\begin{equation}
b \rightarrow 0 \  , \  a \rightarrow 0  \ ,
\end{equation}
with the ratio of $a^2$ and $b$ held fixed,
\begin{equation}
{a^2 \over b} \rightarrow 1  \,\, .
\end{equation}
The result of this contraction is the set of commutation 
relations given in Eq.~(\ref{algebra}).   Lorentz transformations in
the operator algebra are generated by $\hat{M}^{\mu\nu}$, which has the following 
commutation relation with $\hat{\theta}^{\alpha\beta}$:
\begin{eqnarray}
\left[ \hat M^{\mu\nu} , \hat \theta^{\alpha\beta} \right] &=& 
    i  \left( \hat \theta^{\mu\beta} g^{\nu\alpha}
          +   \hat \theta^{\nu\alpha} g^{\mu\beta}
          -   \hat \theta^{\mu\alpha} g^{\nu\beta}
          -   \hat \theta^{\nu\beta} g^{\mu\alpha}     \right) \,\,.
\end{eqnarray}
This is sufficient to establish that $\hat{\theta}^{\mu\nu}$ transforms as
a Lorentz tensor and that Eq.~(\ref{eq:algebra}) is Lorentz covariant.  One may also 
define a momentum operator whose commutation 
relations with $\hat{M}$ and $\hat{\theta}$ are identical to that of $\hat{x}$, 
but this will not be relevant to the subsequent discussion.  Noting that 
$a \rightarrow 0$ is part of the limit, we see that the contracted algebra 
corresponds to a continuum limit of Snyder's quantized spacetime.

With $\hat \theta^{\mu\nu}$ as an additional fundamental operator, 
elements of the group defined locally by Eq.~(\ref{algebra}) depend on 
both $\hat{x}^\mu$ and $\hat{\theta}^{\mu\nu}$.  Ordinary c-number 
functions can again be related to these elements through a Fourier transform, 
though in this case over an extended set of variables. If 
$\hat f = \hat f(\hat x, \hat \theta)$ is a member of the operator
algebra, we define a relation to ordinary functions $f(x,\theta)$ by
\begin{equation}
\hat f = \int (d\alpha) (dB) \,
          e^{-i(\alpha \hat x + B \hat \theta)} \,
          \tilde f(\alpha,B)   \ ,
\label{eq:map1}
\end{equation}
\noindent where $\tilde f$  is the Fourier transform
\begin{equation}
\tilde f(\alpha,B) = \int (dx)(d\theta)   \,
          e^{i(\alpha  x + B  \theta)}  \, f(x, \theta)    \ .
\label{eq:map2}
\end{equation}
In these equations, the measures of integrations are
defined by $(d\alpha) \equiv (2\pi)^{-4} d^4\alpha$, 
$(dB) \equiv (2\pi)^{-6} d^6B$,  $(dx) \equiv d^4x$ and 
$(d\theta) \equiv d^6\theta$; the $B_{\mu\nu}$ and $\theta^{\mu\nu}$
are antisymmetric parameters, and index contraction is implicit in the 
products $\alpha x = \alpha_\mu x^\mu$  and $B\theta \equiv  B_{\mu\nu} 
\theta^{\mu\nu}/2$.   
The measure
\begin{equation}
d^6B = dB_{12} dB_{23} dB_{31} dB_{01} dB_{02} dB_{03}
\end{equation}
\noindent can be shown to be Lorentz invariant if $B_{\mu\nu}$ 
transforms like a second-rank Lorentz tensor.  The $x^\mu$
are a set of ordinary commuting coordinates, and the
$\theta^{\mu\nu}$ (no hat) are a set of new commuting parameters
in ordinary function space that correspond to the $\hat \theta^{\mu\nu}$.  
While the operators $\hat{x}$ and $\hat{\theta}$ are related through 
commutation relations, the commuting parameters $x$ and $\theta$ are 
completely independent of each other. (This reflects the degrees of
freedom associated with the $10$ linearly independent generators of 
SO(4,1).)

The mapping from the operator algebra to the space of ordinary 
functions allows one to define a star-product through the requirement
Eq.~(\ref{eq:thereq}).  The derivation, as usual, begins with the product
\begin{equation}
\hat f \hat g = \int (d\alpha)(dB)(d\gamma)(d\Delta)  \,
          e^{-i(\alpha \hat x + B \hat \theta)} \,
          e^{-i(\gamma \hat x + \Delta \hat \theta)} \,
          \tilde f(\alpha,B)  \,
          \tilde g(\gamma,\Delta)  \ ,
\end{equation}
\noindent which is then simplified using the Baker-Campbell-Hausdorff formula,
\begin{equation}
e^A e^B = e^{A+B + {1\over 2}[A,B] 
       + {1\over 12} [A,[A,B]] + {1\over 12} [B,[B,A]] + \ldots} \ .
\label{eq:bch}
\end{equation}
\noindent As a consequence of Eq.~(\ref{algebra}), the expansion in
Eq.~(\ref{eq:bch}) terminates after the first commutator and, after some 
manipulation, one obtains the same $\star$-product as in the canonical case 
except for the presence of the extra argument $\theta$:
\begin{equation}
(f \star g)(x,\theta) = f(x,\theta)\, \exp[\frac{i}{2} \stackrel{\leftarrow}
{\partial_\mu} \theta^{\mu\nu} \stackrel{\rightarrow}{\partial_\nu}]\,
g(x,\theta) \,\,\, .
\end{equation}
This star product is manifestly Lorentz covariant; the Lorentz transformation 
properties of $\theta$ are identical to those of $\hat{\theta}$, as one can show via 
the mapping defined in Eqs.~(\ref{eq:map1}) and (\ref{eq:map2}).

We also require a generalization of the operator trace.  As 
a trace is a mapping from an operator algebra to numbers that 
is linear, positive (${\rm Tr}\, \hat f \hat f^\dagger \ge 0$), and cyclic
(${\rm Tr}  \hat f \hat g = {\rm Tr} \hat g \hat f$), we propose
\begin{equation}
{\rm Tr} \hat f 
   = \int d^4x \, d^6\theta \, W(\theta) \, f(x,\theta)  \ .
\end{equation}
The weighting function $W(\theta)$ will allow us to work with truncated power series 
expansions of functions in $\theta$.  Therefore, we assume that the weighting function 
is positive  and for any large $|\theta^{\mu\nu}|$ falls to 
zero quickly enough so that all integrals are well defined.  Moreover,
we assume  $W$ is even in  $\theta$, so that
\begin{equation}
\int d^6\theta \, W(\theta) \, \theta^{\mu\nu} = 0  \ .
\end{equation}
Field theory actions follow from Eq.~(\ref{eq:actdef}),
\begin{equation}
S = \int d^4x \, d^6\theta \, W(\theta)\, {\cal L}(\phi,\partial\phi)_\star \,\,\, .
\end{equation}
As ${\cal L}(\phi,\partial\phi)_\star$ depends in general on both $x$ and $\theta$, 
the object that takes the role the ordinary Lagrangian will be the 
$\theta$-integrated quantity,
\begin{equation}
{\cal L}(x) = \int d^6\theta \, W(\theta) \, {\cal L}(\phi,\partial\phi)_\star \ .
\end{equation}


\section{Gauge Theory}\label{sec:gauge}


The star product that we have formulated allows us to reproduce the
noncommutativity of the operators $\hat{x}^\mu$ and $\hat\theta^{\mu\nu}$
while working instead with functions of the classical variables $x^\mu$ 
and $\theta^{\mu\nu}$.  Ordinary quantum field theories involve fields 
that are functions of $x$ alone, suggesting two possible ways to proceed.
For a theory without gauge invariance, we may simply choose our fields 
$\phi(x,\theta)$ to be functions of $x$ only
\begin{equation}
\phi(x,\theta) \equiv \phi(x)  \,\,\, ,
\end{equation}
and construct an action using the trace described in the previous section.  
For example, the Lagrangian for $\phi^4$ theory is
\begin{equation}
{\cal L} =  \frac{1}{2} \partial_\mu \phi 
\partial^\mu \phi-\frac{1}{2}m^2 \phi^2 - \frac{\lambda}{4\!} 
\int d^6 \theta \, W(\theta)\,  (\phi\star\phi)^2 
\,\,\, .
\end{equation}
Here we have used 
\begin{equation}
\int d^4x \, f \star g = \int d^4x \, f \, g  \,\,\, ,
\end{equation}
and the normalization condition
\begin{equation}
\int d^6\theta \, W(\theta) = 1 \,\, ,
\end{equation}
to simplify the result.

On the other hand, if the field $\phi$ transforms as some representation of 
a gauge group $G$, then it is no longer possible to choose $\phi$ to be a 
function of $x$ only, as $\theta$ dependence is introduced via the 
noncommutative generalization of the gauge transformation.  Consider a U(1)
gauge theory with a matter field $\psi$ and gauge field $A$.  Under a gauge
transformation parameterized by $\Lambda(x,\theta)$, the fields transform 
as
\begin{equation}
\psi(x,\theta) \rightarrow \psi'(x,\theta) = U \star 
\psi(x,\theta) \,\,\, ,
\label{eq:psinc}
\end{equation}
\begin{equation}
A_\mu(x, \theta) \rightarrow A'_\mu(x, \theta)= U\star 
A_\mu(x, \theta) \star U^{-1} + \frac{i}{e}\, U \star \partial_\mu U^{-1} 
\,\,\, ,
\label{eq:anc}
\end{equation}
where
\begin{equation}
U = (e^{i \Lambda})_\star  \,\,\, .
\end{equation}
It is straightforward to confirm that the Lagrangian
\begin{equation}
{\cal L} = \int d^6 \theta \, W(\theta) \left[
-\frac{1}{4} F_{\mu\nu} \star F^{\mu\nu}
+ \bar{\psi}\star (i\not\!\!D-m)\star \psi\right]
\label{eq:newqed}
\end{equation}
is gauge invariant, provided that
\begin{equation}
D_\mu=\partial_\mu-i e A_\mu  \,\,\, ,
\end{equation}
and
\begin{equation}
F_{\mu\nu}=\partial_\mu A_\nu - \partial_\nu A_\mu - 
i e [A_\mu\stackrel{\star}{\,,}A_\nu] \, .
\label{eq:fmunu}
\end{equation}
Superficially, Eqs.~(\ref{eq:newqed}--\ref{eq:fmunu}) are the same 
as in the case of canonical noncommutative QED~\cite{HAY}, aside from 
the fact that our construction of the trace averages over the parameter 
$\theta$.  One must keep in mind, however, that the fields in 
Eqs.~(\ref{eq:newqed}--\ref{eq:fmunu}) are functions of both $x$ 
and $\theta$, and cannot be identified with the ordinary quantum fields 
$\psi(x)$ and $A^\mu(x)$.

To proceed, we will expand the fields as a power series in 
the variable $\theta$, and demonstrate that the coefficients, which are 
functions of $x$ alone, can be expressed solely in terms of 
ordinary quantum fields.  The  nonlinear field redefinition is fixed by 
the constraints of noncommutativity and gauge invariance. This approach is
largely the same as the one employed in the construction of SU(N) 
noncommutative gauge theories in Refs.~\cite{Jurco,Calmet}.  The expansion 
in $\theta$ in our case is valid given the presence of the weighting 
function $W(\theta)$ that renders the integral of higher order terms small.  
Let us demonstrate this approach by constructing the Lagrangian for the pure 
gauge sector of our U(1) theory.

We begin by expanding both the gauge parameter $\Lambda$ and the gauge field
$A^\mu$ 
\begin{equation}
\Lambda_\alpha(x,\theta)=\alpha(x) + \theta^{\mu\nu} \Lambda_{\mu\nu}^{(1)}(x;\alpha) + 
\theta^{\mu\nu}\theta^{\eta\sigma}\Lambda_{\mu\nu\eta\sigma}^{(2)}(x;\alpha)
+\cdots \,\,\, ,
\label{eq:lamexpand}
\end{equation}
\begin{equation}
A_\rho(x,\theta)= A_\rho(x) + \theta^{\mu\nu} A^{(1)}_{\mu\nu\rho}(x) + 
\theta^{\mu\nu}\theta^{\eta\sigma} A^{(2)}_{\mu\nu\eta\sigma\rho}(x)+\cdots 
\,\,\, .
\label{eq:aexpand}
\end{equation}
We identify the first term in each expansion as the ordinary, $x$-dependent 
gauge parameter and gauge field, respectively.  In an Abelian gauge theory, 
two gauge transformations parameterized by $\alpha(x)$ and $\beta(x)$ 
satisfy the property that
\begin{equation}
(\delta_\alpha\delta_\beta -\delta_\beta\delta_\alpha) \psi(x) =0  \,\,\, ,
\label{eq:abelian}
\end{equation}
where $\psi$ is a matter field transforming infinitesimally as
\begin{equation}
\delta_\alpha \psi(x) = i \, \alpha(x) \psi(x)  \,\,\, .
\end{equation}
In the noncommutative theory, we require that the field $\psi(x,\theta)$ 
satisfy
\begin{equation}
(\delta_\alpha\delta_\beta -\delta_\beta\delta_\alpha) \psi(x,\theta) =0  
\,\,\, ,
\label{eq:ncabelian}
\end{equation}
where the infinitesimal transformation 
\begin{equation}
\delta_\alpha \psi(x,\theta) = i \, \Lambda_\alpha(x,\theta) \star \psi(x,\theta) 
\label{eq:ncin}
\end{equation}
follows from Eq.~(\ref{eq:psinc}).  Eq.~(\ref{eq:ncabelian}) requires that 
the parameter
$\Lambda$ satisfy
\begin{equation}
i \delta_\alpha \Lambda_\beta-i\delta_\beta \Lambda_\alpha + [\Lambda_\alpha 
\stackrel{\star}{\, ,}
\Lambda_\beta]=0  \,\, ,
\end{equation}
from which the transformation properties of $\Lambda^{(1)}$ and 
$\Lambda^{(2)}$ in 
Eq.~(\ref{eq:lamexpand}) may be deduced.  It may then be shown~\cite{Jurco} that the 
following functions of the ordinary gauge parameter and gauge field satisfy 
this consistency condition:
\begin{equation}
\Lambda_{\mu\nu}^{(1)}(x;\alpha) = \frac{e}{2} \partial_\mu \alpha(x) A_\nu(x) 
\,\,\, ,
\end{equation}
\begin{equation}
\Lambda_{\mu\nu\eta\sigma}^{(2)}(x;\alpha) = -\frac{e^2}{2} \partial_\mu \alpha(x) 
A_\eta(x) \partial_\sigma
A_\nu(x)  \,\,\, .
\end{equation}
Similarly, the requirement that the noncommutative gauge field $A(x,\theta)$ 
transforms
infinitesimally as
\begin{equation}
\delta_\alpha A_\sigma = \partial_\sigma \Lambda_\alpha 
+ i [\Lambda_\alpha \stackrel{\star}{\,,} A_\sigma] \,\,\, ,
\end{equation}
which follows from Eq.~(\ref{eq:anc}), is sufficient to determine the 
correct transformation properties of $A^{(1)}$ and $A^{(2)}$.  These 
are reproduced by 
\begin{equation}  
A^{(1)}_{\mu\nu\rho}(x)=-\frac{e}{2}A_\mu(\partial_\nu A_\rho 
+ F^0_{\nu\rho})  \,\,\, ,
\end{equation}
\begin{equation}
A^{(2)}_{\mu\nu\eta\sigma\rho}(x)=
\frac{e^2}{2}(A_\mu A_\eta \partial_\sigma F^0_{\nu\rho}
-\partial_\nu A_\rho\partial_\eta A_\mu A_\sigma
+ A_\mu F^0_{\nu\eta} F^0_{\sigma\rho})  \,\,\, ,
\end{equation}
where $F^0_{\mu\nu}$ represents the ordinary Abelian field strength tensor
\begin{equation}
F^0_{\mu\nu}=\partial_\mu A_\nu-\partial_\nu A_\mu  \,\,\, .
\end{equation}
We may now express the noncommutative field strength tensor in terms of the 
ordinary gauge field $A^\mu(x)$.  We find
\begin{eqnarray}
F_{\mu\nu}&=& F^0_{\mu\nu}+ e\, \theta^{\kappa\lambda}(F^0_{\mu \kappa}
F^0_{\nu \lambda}-
A_\kappa \partial_\lambda F^0_{\mu\nu})
+\frac{e^2}{2} \theta^{\kappa\lambda}\theta^{\rho\eta} [ F^0_{\kappa\nu} 
F^0_{\lambda \rho} 
F^0_{\eta\mu}
-F^0_{\kappa\mu} F^0_{\lambda \rho} F^0_{\eta \nu} \nonumber \\
&+& A_\kappa ( 2 F^0_{\mu \rho} \partial_\lambda F^0_{\eta \nu} 
- 2 F^0_{\nu \rho} \partial_\lambda F^0_{\eta \mu}+F^0_{\lambda \rho} 
\partial_\eta F^0_{\mu\nu}
- \partial_\rho F^0_{\mu\nu} \partial_\lambda A_\eta) 
+ A_\kappa A_\rho \partial_\eta \partial_\lambda 
F^0_{\mu\nu}] .
\end{eqnarray}
Photon self-interactions may be isolated by substituting this result into 
Eq.~(\ref{eq:newqed}) and integrating over $\theta$.  For any 
Lorentz-invariant weighting function 
$W(\theta)\equiv W(\theta_{\mu\nu} \theta^{\mu\nu})$,
\begin{equation}
\int d^6\theta \, W(\theta) \, \theta^{\mu\nu} \theta^{\eta\rho}
=\frac{1}{12} \langle \theta^2 \rangle (g^{\mu\eta}g^{\nu\rho}-
g^{\mu\rho} g^{\eta\nu})  \,\, ,
\label{eq:trick}
\end{equation}
where we have defined the expectation value
\begin{equation}
\langle \theta^2 \rangle \equiv \int d^6 \theta \, W(\theta) 
\, \theta_{\mu\nu} \theta^{\mu\nu} \,\,\, .
\end{equation}
Photon self-interaction terms that are odd in $\theta$ vanish under this 
integration, so that the lowest-order nonstandard vertex is given by:
\begin{equation}
{\cal L} = \frac{\pi\alpha}{12} \langle \theta^2 \rangle [ F^0_{\mu\nu} F^{0\nu\eta} 
F^0_{\eta\rho} F^{0\rho\mu}
- (F^0_{\mu\nu} F^{0\mu\nu})^2]  \,\,\, .
\label{eq:result}
\end{equation}
Notice that the $ \langle \theta^2 \rangle ^{-1/4}$ is an energy scale that 
characterizes the onset of new physics; since one generally likes to 
keep this scale a free parameter in phenomenological studies, one need 
not specify anything more about the form of the weighting function 
$W(\theta)$, at least at the order to which we are working.  It is 
interesting to note that Eq.~(\ref{eq:result}) reduces to
\begin{equation}
{\cal L}=  \frac{\pi\alpha}{6} \langle \theta^2 \rangle [- ({\bf E}^2-{\bf B}^2)^2 
+ 2 ({\bf E} \cdot 
{\bf B})^2]
\end{equation}
when expressed in terms of classical electric and magnetic fields, which 
differs in form from the famous Euler-Heisenberg low-energy 
effective Lagrangian following at the 
one-loop level in QED~\cite{euler} 
\begin{equation}
{\cal L}_{E-L}= \frac{2 \alpha^2}{45 m_e^4} \, [({\bf E}^2-{\bf B}^2)^2
+ 7 ({\bf E} \cdot {\bf B})^2] 
\,\,\, .
\label{eq:elag}
\end{equation}
Eq.~(\ref{eq:elag}) is valid at energies small compared to the electron mass,
while our expectation is that $\langle \theta^2 \rangle^{-1/4}$ will be of order a 
TeV, based on the type of bounds that are typical in extensions of the standard model 
that modify the gauge sector.  We therefore turn briefly to the high-energy collider 
physics of our scenario, where the effects of noncommutativity are more likely to be 
manifest.


\section{Phenomenology}\label{sec:phenom}


The vertices that follow from our Lorentz-invariant construction of 
noncommutative QED provide a rich hunting ground for the origins of new 
phenomena at colliders.  Deviations in observable scattering cross sections 
follow from modifications to vertices that occur at tree level in the 
standard model, as well as from the existence of new vertices.  Examples of 
the latter include direct two-photon-two-fermion couplings, as well as the 
four-photon interaction discussed in detail in the previous section.  Here 
will focus on the scattering process $\gamma\gamma\rightarrow \gamma\gamma$, 
which has been discussed in the recent literature as a potential window on 
physics beyond the standard model~\cite{gounaris,gggglit}.   

\begin{figure}[t]
\centerline {\epsfxsize 2. in \epsfbox{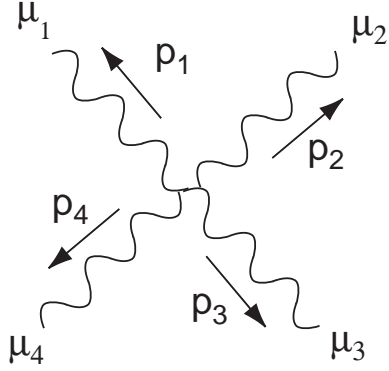}}
\caption{Four-photon vertex. \label{fig:fd}}
\end{figure}
Given the labeling of momenta and Lorentz indices shown in Fig.~\ref{fig:fd},
the interaction in Eq.~(\ref{eq:result}) leads to the Feynman rule
\begin{eqnarray}
V_{4\gamma} &=& {1\over 6} i e^2 \left\langle \theta^2 \right\rangle
\Bigg\{ -  4 p_1^{\mu_2} p_2^{\mu_1} p_3^{\mu_4} p_4^{\mu_3}
       +    p_1^{\mu_2} p_2^{\mu_3} p_3^{\mu_4} p_4^{\mu_1}
       +    p_1^{\mu_2} p_2^{\mu_4} p_3^{\mu_1} p_4^{\mu_3}
                          \nonumber   \\ &&\hskip 5.25em
       -  4 p_1^{\mu_3} p_2^{\mu_4} p_3^{\mu_1} p_4^{\mu_2}
       +    p_1^{\mu_3} p_2^{\mu_1} p_3^{\mu_4} p_4^{\mu_2}
       +    p_1^{\mu_3} p_2^{\mu_4} p_3^{\mu_2} p_4^{\mu_1}
                          \nonumber   \\ &&\hskip 5.25em
       -  4 p_1^{\mu_4} p_2^{\mu_3} p_3^{\mu_2} p_4^{\mu_1}
       +    p_1^{\mu_4} p_2^{\mu_1} p_3^{\mu_2} p_4^{\mu_3}
       +    p_1^{\mu_4} p_2^{\mu_3} p_3^{\mu_1} p_4^{\mu_2}
                          \nonumber   \\
&+&   \Big( g^{\mu_1 \mu_2} 
      \bigg[ \ (p_1^{\mu_3} p_2^{\mu_4} + p_2^{\mu_3} p_1^{\mu_4})
                   \,  p_3 \cdot p_4
       +  4 p_4^{\mu_3} p_3^{\mu_4}  p_1 \cdot p_2
                          \nonumber   \\ &&\hskip 2.25em
       -    p_4^{\mu_3} p_1^{\mu_4}  p_2 \cdot p_3
       -    p_4^{\mu_3} p_2^{\mu_4}  p_1 \cdot p_3
       -    p_1^{\mu_3} p_3^{\mu_4}  p_2 \cdot p_4
       -    p_2^{\mu_3} p_3^{\mu_4}  p_1 \cdot p_4   \bigg]
                          \nonumber   \\
&&+  \left[ \ (12)(34) \rightarrow (34)(12)  \ \right]
                          \nonumber   \\
&&+  \left[ \ (12)(34) \rightarrow (13)(42)  \ \right]  +
     \left[ \ (12)(34) \rightarrow (42)(13)  \ \right]
                          \nonumber   \\
&&+  \left[ \ (12)(34) \rightarrow (14)(23)  \ \right]  +
     \left[ \ (12)(34) \rightarrow (23)(14)  \ \right]   \Big)
                          \nonumber   \\
&+&  \Big(   g^{\mu_1 \mu_2} g^{\mu_3 \mu_4}
     \bigg[ - 4  p_1 \cdot p_2 \  p_3 \cdot p_4
            +    p_1 \cdot p_4 \  p_2 \cdot p_3
            +    p_1 \cdot p_3 \  p_2 \cdot p_4  \bigg]
                          \nonumber   \\
&&+  \left[ \ (12)(34) \rightarrow (13)(42)  \ \right]  +
     \left[ \ (12)(34) \rightarrow (14)(23)  \ \right] \Big) \Bigg\}  \,\, .
\end{eqnarray}

Placing the photons on shell, we may compute the differential scattering 
cross section in the photon center-of-mass frame.  The noncommutative amplitude
is $90^\circ$ out of phase with the leading logarithmic contributions to the
standard model background (see below), so we may write
\begin{equation}
\sigma \approx \sigma_{{\rm NC}} + \sigma_{{\rm SM}}  \,\,\, .
\end{equation}
For unpolarized beams, we find
\begin{equation}
\frac{d\sigma_{{\rm NC}}}{d\cos\Theta^*} = \frac{19\pi}{128} \left(\frac{\langle 
\theta^2 \rangle}{12}\right)^2 \alpha^2 s^3 
(3+\cos^2\Theta^*)^2 \,\,\, ,
\end{equation}
where $\sqrt{s}$ and $\Theta^*$ are the center of mass energy and scattering 
angle, respectively.  It then follows that the noncommutative contribution to the 
total cross section ($0^\circ < \Theta^*<180^\circ$) is given by
\begin{equation}
\sigma_{{\rm NC}} = \frac{133\pi}{80} \alpha^2 s^3 \left(\frac{\langle \theta^2 \rangle}
{12}\right)^2  \,\,\, .
\end{equation}

\begin{figure}[t]
\centerline {\epsfxsize 4. in \epsfbox{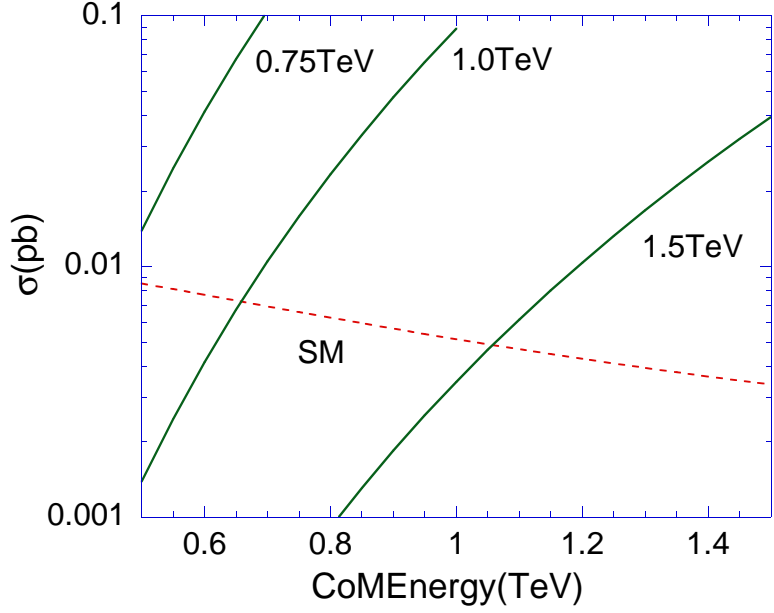}}
\caption{Total cross sections $\sigma_{{\rm NC}}$ and $\sigma_{{\rm SM}}$
for $30^\circ < \Theta^* < 150^\circ$.  Noncommutative results are labeled by
the value of $\Lambda_{{\rm NC}}$, defined in the text.}\label{fig:totxs}
\end{figure} 
To compare our result to the expectation in the standard model, we use the 
amplitudes given in Ref.~\cite{gounaris} for light-by-light scattering in 
the high-energy limit $s,\, |t|,\,|u| \gg m_W^2$.  So that our discussion 
is self-contained, we reproduce the relevant results.  The differential 
cross section is given by
\begin{equation}
\left(\frac{d\sigma}{d\cos\Theta^*}\right)_{SM} = \frac{1}{128\pi s} 
[ (\mbox{Im } F_{++++})^2+
(\mbox{Im } F_{+-+-})^2 + (\mbox{Im } F_{+--+})^2]  \,\,\, ,
\end{equation}
where the dominant helicity amplitudes are mostly imaginary and 
\begin{equation}
 \mbox{Im } F_{++++} = -16 \pi \alpha^2 \left[ \frac{s}{u} \ln \Big| 
\frac{u}{m_W^2}\Big|
+\frac{s}{t} \ln \Big| \frac{t}{m_W^2} \Big| \right]  \,\,\, ,
\end{equation}
\begin{equation}
\mbox{Im } F_{+-+-} = -12 \pi \alpha^2 \frac{s-t}{u}-16 \pi \alpha^2 \left[
\frac{u}{s} \ln \Big| \frac{u}{m_W^2} \Big|
+\frac{u^2}{s t} \ln \Big| \frac{t}{m_W^2} \Big| \right] \,\,\, ,
\end{equation}
with
\begin{equation}
\mbox{Im }F_{+--+}(s,t,u) = \mbox{Im }F_{+-+-}(s,u,t) \,\,\, .
\end{equation}

\begin{figure}
\centerline {\epsfxsize 4. in \epsfbox{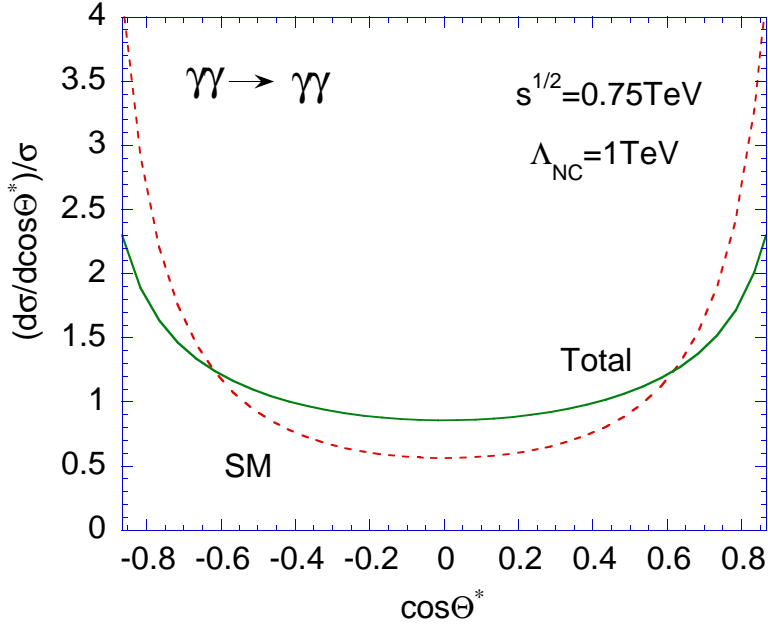}   }
\caption{Differential cross sections for $\sqrt{s}=0.75$~TeV and 
$\Lambda_{{\rm NC}}=1.0$~TeV, normalized to $\sigma (30^\circ < \Theta^* < 150^\circ)$. 
The dashed line indicates the standard model background and the solid line indicates 
the result when both the standard model and Lorentz-invariant NCQED interactions are 
present.}\label{fig:difxs}
\end{figure}
Figs.~\ref{fig:totxs} and \ref{fig:difxs} show the comparison between our
noncommutative result and the expectation in the standard model.  Since the scale
of new physics $\Lambda_{{\rm NC}}$ is characterized by a root-mean-square
average of the components of $\theta^{\mu\nu}$, we define
\begin{equation}
\Lambda_{NC} = \left(\frac{12}{\langle \theta^2 \rangle}\right)^{1/4} \,\,\, ,
\label{eq:lnc}
\end{equation}
which also is a natural choice given Eq.~(\ref{eq:trick}).  Note that the
effective expansion parameter in the scattering amplitude is 
$s^2 \langle \theta^2 \rangle /12 \equiv s^2/\Lambda_{{\rm NC}}^4$, and each curve in 
Fig.~\ref{fig:totxs} falls within a range where this ratio is less than one. The 
reader may easily estimate the size of higher-order corrections at any point in 
Fig.~\ref{fig:totxs} by computing $s^2/\Lambda_{{\rm NC}}^4$. While the total cross 
section rises as $s^3$, which one would expect generically given the presence of new, 
effective contact interactions, the angular distribution is
less forward and backward peaked in comparison to the standard model result. 
From the effective field theory point of view, any new physics can be 
parameterized by gauge-invariant interactions of the form 
$c_1 F_{\mu\nu} F^{\nu\eta} F_{\eta\rho} F^{\rho\mu} 
+ c_2 (F_{\mu\nu} F^{\mu\nu})^2$, for some coefficients $c_1$ and
$c_2$.  (Other possible interactions involving derivatives are irrelevant for
a process in which all the photons are on shell.) While the scaling of the
cross section with energy follows simply from dimensional analysis, the precise
form of the dependence on scattering angle depends on the relative values of these 
coefficients. Note that our plots are evaluated for and 
$30^\circ < \Theta^* < 150^\circ$, the same angular range adopted in 
Ref.~\cite{gounaris},  which eliminates events close to the beam direction.
For this choice, there are points in Fig.~\ref{fig:totxs} where the
noncommutative cross section substantially exceeds the standard model result, 
higher-order corrections in $\theta$ are under control, and our initial 
kinematical assumptions are satisfied. In a more complete phenomenological study, 
one would take
into account the energy distribution of the initial 
photons, which are not monochromatic when produced via laser backscattering 
at an $e^+e^-$ linear collider like CLIC or the NLC.  Moreover, one can 
extract additional information from the polarized cross section since the 
polarization of the incident photon beams can be controlled to a large 
extent by the polarization of the lepton beams.  We hope it is clear from 
the present example that our scenario may lead to potentially distinctive 
collider signals, and defer a complete investigation of these phenomenological 
issues to future work.


\section{Conclusions}\label{sec:conc}

We have formulated a new class of noncommutative field theories 
in which the coordinate commutation relations are Lorentz covariant:
\begin{equation}
\left[ \hat x^\mu , \hat x^\nu \right] = i \hat \theta^{\mu\nu}  \,\,\, .
\label{eq:again}
\end{equation}
Here the parameter $\theta^{\mu\nu}$ of canonical noncommutative theories
has been promoted to an operator $\hat{\theta}^{\mu\nu}$ that transform 
like a Lorentz tensor and all other relevant commutators are vanishing.
We showed how Eq.~(\ref{eq:again}) may be realized through the contraction of
a larger Lorentz-invariant algebra for which explicit representations are 
already known.

Functions in the algebra of Eq.~(\ref{eq:again}) depend on both $\hat{x}$ 
and $\hat{\theta}$.  We may map these to functions of commuting variables provided 
the rule for multiplication is modified.  We presented the star product of functions
$f(x,\theta)$ that mimics the multiplication of operator functions $\hat{f}(\hat{x},
\hat{\theta})$.  By necessity, the commuting functions may depend not only
on the familiar commuting variables $x^\mu$, but also on a new set $\theta^{\mu\nu}$, 
that we treat as unphysical parameters; the operator trace may be expressed as an
integral over both $x$ and $\theta$.  With a star product and trace at hand, we 
showed how to formulate field theories in terms of functions of $x^\mu$ alone, 
and how to maintain gauge invariance through nonlinear field redefinitions. 

We applied our formalism in constructing a Lorentz-invariant version of noncommutative
QED.  New vertices are present in this theory that are not found in ordinary QED, 
including two-fermion-two-photon and four-photon interactions, to name a few. 
However, unlike canonical noncommutative QED, no three-photon vertex is present. 
As an example of what might be observed experimentally if Lorentz-invariant 
noncommutative QED describes nature, we considered photon-photon elastic scattering 
at high energies, and obtained contributions that are significant with respect to the 
standard model background. The new noncommutative amplitude is present at tree level 
and at lower order in $e^2$ than the one-loop standard model result. The scattering 
cross section was shown to differ in both its energy dependence and angular 
distribution.  At a photon-photon collider with $\sqrt{s}=500$~GeV and an annual 
integrated luminosity of 100~fb$^{-1}$, one expects thousands of standard model 
events, while for $\Lambda_{{\rm NC}}=0.75$~TeV the noncommutative effects can yield 
O($100$\%) corrections.  Our results suggests that there is a clear opportunity at 
colliders to see the effects of Lorentz-conserving noncommutative QED if the 
noncommutativity scale is on the order of a TeV.

%
\begin{acknowledgments}
C.E.C. and C.D.C  thank the National Science Foundation for support under 
Grant No.\ PHY-9900657.  In addition, C.D.C. thanks the Jeffress Memorial 
Trust for support under Grant No.~J-532.
\end{acknowledgments}


\end{document}